\documentclass[showpacs,preprintnumbers, twocolumn,
amsmath,amssymb,APSl,prd,nofootinbib,superscriptaddress]{revtex4-2} 
\usepackage{graphicx}
\usepackage{amsmath,amsthm,amssymb}
\usepackage{hyperref}
\usepackage{xcolor}
 
\hypersetup{colorlinks=true, linkcolor=blue, citecolor=green}

 \newcommand{\be}{\begin{equation}}
\newcommand{\ee}{\end{equation}}
\newcommand{\bea}{\begin{eqnarray}}
\newcommand{\eea}{\end{eqnarray}}
\newcommand{\beaa}{\begin{eqnarray*}}
\newcommand{\eeaa}{\end{eqnarray*}}

\begin{document}
 
 
\title{Light trajectories and optical appearances in asymptotically Anti-de Sitter-Schwarzschild and black string space-times}
\author{G. Alencar}
\email{geova@fisica.ufc.br}
\affiliation{Department of Physics, Universidade Federal do Cear\'a (UFC), Campus do Pici, Fortaleza - CE, C.P. 6030, 60455-760 - Brazil}

\author{Arthur Lima}
\email{arthur.lima@fisica.ufc.br}
\affiliation{Department of Physics, Universidade Federal do Cear\'a (UFC), Campus do Pici, Fortaleza - CE, C.P. 6030, 60455-760 - Brazil}

\author{Diego Rubiera-Garcia} \email{drubiera@ucm.es}
\affiliation{Departamento de F\'isica Te\'orica and IPARCOS,
	Universidad Complutense de Madrid, E-28040 Madrid, Spain}
    
\author{Diego~S\'aez-Chill\'on~G\'omez}
\email{diego.saez@uva.es}
\affiliation{Department of Theoretical, Atomic and Optical
Physics, and Laboratory for Disruptive Interdisciplinary Science (LaDIS), Campus Miguel Delibes, \\ University of Valladolid UVA, Paseo Bel\'en, 7, 47011
Valladolid, Spain}
\affiliation{Department of Physics, Universidade Federal do Cear\'a (UFC), Campus do Pici, Fortaleza - CE, C.P. 6030, 60455-760 - Brazil}

\begin{abstract}

The Event Horizon Telescope (EHT) imaging of the central objects in the M87 and Milky Way galaxies provide compelling evidence that these objects are consistent with (Kerr) black holes. In view of these observations and the future expectations of Very Long Baseline Interferometry (VLBI) on which the EHT observations are based, an intensive research work has been carried out in the literature to simulating light trajectories and reconstructing the corresponding optical appearance for a wide array of modified black holes and ultra-compact objects. The corresponding images are directly affected not only by the background space-time geometry but also by the physics of the accretion disk, whose combination yields a characteristic fingerprint. In this paper, we consider such a fingerprint for objects which are not asymptotically flat but instead approach a Anti-de Sitter space-time. This assumption significantly influences light trajectories and, consequently, the corresponding images of the objects as seen by an observer at some distance, which can be used in future VLBI observations for testing alternatives of this kind to the Kerr paradigm.  We illustrate our considerations with the examples of a Schwarzschild-Anti-de Sitter black hole and a black string, discussing their most notable departures from canonical, asymptotically-flat black hole space-times.

\end{abstract}

\maketitle

\section{Introduction}
\label{Intro}

The images obtained by the Event Horizon Telescope (EHT) Collaboration of the supermassive objects located at the center of the M87 and Milky Way galaxies have opened a new era for the study of the strong-field regime in gravitational physics \cite{EventHorizonTelescope:2019dse,EventHorizonTelescope:2022wkp}. Such images display a bright ring-like region surrounding a central brightness depression, in agreement with the results of General Relativistic MagnetoHydroDynamical (GRMDS) simulations of the accretion flow \cite{Chan:2014nsa}. The finer details of such features are, however, determined by a complex interplay between the accretion disk physics and the background space-time geometry \cite{Younsi:2021dxe}. The latter determines the existence (provided the object is compact enough, which includes black holes but also other objects) of a key ingredient in the generation of images, namely, the photon shell (the photon sphere in the spherically symmetric case), corresponding to the surface of (unstable) bound photon orbits above the event horizon. Light trajectories passing nearby the photon shell suffer strong deflections \cite{Perlick:2021aok}, allowing for several turns around the compact object and being at the heart of both bright ring and shadow features in the corresponding images.

Within General Relativity (GR), the features of the photon shell rely on the Kerr family of solutions, the unique vacuum, asymptotically flat, axi-symmetric space-time solution of Einstein's field equations, and characterized by mass and angular momentum \cite{Kerr:1963ud}. Such a family provides clear predictions for astrophysical phenomena, in particular, in terms of its compatibility of its cast images with the observed ones, in turn becoming a cornerstone for tests of GR itself \cite{Mizuno:2018lxz,Paugnat:2022qzy,EventHorizonTelescope:2020qrl,Gralla:2019xty}. However, the complex task of extracting useful, detailed information from such images for the sake of testing the exact description provided by the  Kerr metric has still a large path ahead. Two main challenges are present: the difficulty to separate astrophysics and geometry in their respective contributions to the images \cite{Lara:2021zth}, and the current limited capabilities to resolving tiny interferometric signatures \cite{Johnson:2019ljv}. Regarding the first challenge, the physics of the accretion disk is still not well understood in terms of its geometrical, optical and emission features requiring a heavy modelling \cite{Gold:2020iql}, but also regarding magnetic and polarization features of the accreting material \cite{EventHorizonTelescope:2025vum} and furthermore in the effects enhanced by any surrounding matter in the observer's line of sight \cite{Xavier:2023exm}. In addition, alternative gravitational configurations, both in terms of modified or hairy black holes \cite{Herdeiro:2015waa} and in ultra-compact, horizonless compact objects \cite{Cardoso:2019rvt} can also mimic the features of black hole images (see e.g. \cite{Olivares:2018abq,Vincent:2020dij} for a discussion). As for the second challenge, future VLBI projects such as the ngEHT or the Black Hole Explorer  \cite{Tiede:2022grp,Lupsasca:2024xhq}, are expected to provide the required capability to resolve the finer structure hidden in the bright ring, namely, its {\it photon rings}, which encodes most of the promises for testing gravitational physics within this context. Moreover, the existing overlapping of photon rings might be used to constrain the spacetime metric as well \cite{Aratore:2024bro,Tsupko:2025hhf}.

Motivated by these developments, a great deal of effort has been invested in the community to simulating the optical appearance of alternative objects to the Kerr solution, such as modified \cite{Amarilla:2010zq,Atamurotov:2013dpa,Wang:2018prk,Vagnozzi:2023qyf,Cui:2024wvz} and hairy black holes \cite{Vincent:2016sjq,Sengo:2022jif}, including parametrized black hole solutions \cite{Younsi:2016azx,Olmo:2025ctf}, but also horizonless, ultra-compact objects and black hole mimickers  of different sorts \cite{Vincent:2015xta,Simpson:2018tsi,Bacchini:2021fig,Peng:2021wrs,Huang:2023wrt,Saurabh:2023otl,Guo:2022iiy,Wang:2023nwd,KumarWalia:2024yxn,Chen:2024ibc,He:2025qmq,deSa:2025nsx}. Alternatives to the Kerr black hole (Schwarzschild in the spherically symmetric case) may modify the number, type, and location of the photon sphere, which affects directly the reconstructed image as observed from far away observers, most notably on its structure of photon rings and shadows \cite{daSilva:2023jxa}. To achieve this end, these approaches face technical, theoretical, numerical, and phenomenological difficulties to extract clear and clean observables brushing off the contamination enhanced by the uncertainties in the modelling of the accretion disk. The outcome of this collective effort is to ready ourselves for a confrontation of the predictions on the optical appearance of compact objects with foreseeable observations and, ultimately, to provide fundamental information on the nature of compact objects, the structure of the space-time around it, and GR itself \cite{Younsi:2023review}. 





The main aim of this work is to study the optical appearances of black hole space-times  immersed in asymptotically Anti-de Sitter (AdS) space-times, i.e. those with the presence of a negative cosmological constant.  Previous works in the literature have analyzed the case of asymptotically de Sitter space-times \cite{Wang:2023rjl,Cao:2024kht}, which is a more accurate description of the accelerating cosmological expansion of the universe according to observational data \cite{Huterer:2017buf}. However, AdS space-times are interesting on their own through its well known correspondence with Conformal Field Theories (CFTs) \cite{Aharony:1999ti}, while  black holes in such spaces have been previously studied in the literature \cite{Cvetic:2001bk,Cai:2004eh,Wang:2014eha,Xu:2017bpz,Ashraf:2025uhj}. Furthermore, the analysis of asymptotically AdS space-times provides understanding on the  reconstruction of the images of non-asymptotically flat space-times via modifications to the location of bound orbits and, consequently, to the photon rings structure as viewed by an observer at certain distance. We shall illustrate our discussion with two such space-times:  one that describes a Schwarzschild-AdS black hole \cite{AdSBlackHoles}, and another one describing a black string, namely, a black hole with cylindrical symmetry in an asymptotically AdS  with a $\mathcal{R} \times \mathcal{S}^1$ topology \cite{Lemos:1994xp,Bronnikov:2009na}.

The paper is organized as follows: in Sec. \ref{AdSspacetimes} we briefly review  the main features of asymptotically Anti-de Sitter space-times, focusing on the two main geometries considered in this paper. In Sec. \ref{Geo} null geodesics and the ray-tracing procedure are applied to such space-times. Sec. \ref{MPIE} deals with the corresponding images by a geometrically and optically thin accretion disk, using a set of emission profiles with peaks at different locations, and for various inclinations of the disk. Finally, Sec. \ref{Conclusions} gathers the conclusions of the paper.

\section{Asymptotically Anti-de Sitter spacetimes}
\label{AdSspacetimes}

The kind of space-times we are interested in this work are those which approach an AdS metric at asymptotic infinity. We shall thus start by reviewing the $4-$dimensional AdS space-time, whose line element is given by:
\be
ds^2=A(r) dt^2+ A^{-1}(r) dr^2+r^2d\Omega^2\ ,
\label{GeneralMetric}
\ee
where 
\be
 A(r)=1+\alpha^2r^2\ ,
 \label{AdSspacetime}
\ee
with $\alpha^2=-\frac{\Lambda}{3}$ and the cosmological constant being $\Lambda<0$. This space-time is an embedding of an hyperboloid in five dimensions, whose (periodic) time coordinate is unrolled in order to avoid closed time-like curves. Since it is conformally equivalent to half of the Einstein static universe, a time-like boundary on the space-time exists, which can be reached by massless particles but massive particles are not able to, since its effective potential grows with the square of the distance. Despite asymptotically AdS space-times are generally not believed to describe realistic physical systems, their analysis has been always considered as of great importance, specially concerning some quantum gravity aspects and in view of the AdS/CFT correspondence. In the following we review two asymptotically AdS space-times whose cast images are the main goal of this paper.

\subsection{Schwarzschild-AdS black hole}

\begin{figure}[t!]
\centering
\includegraphics[scale=0.66]{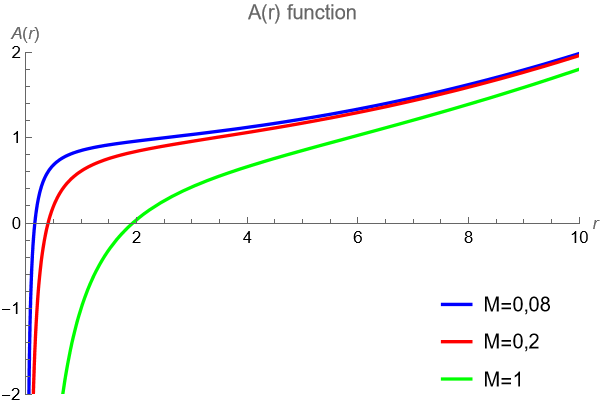}
\caption{The metric function $A(r)$ for different values of the $M$ parameter  for the metric of Schwarzschild-AdS black hole (\ref{SAdSMetric}) with
 $\alpha^2=0.1$.}
\label{Fig3}
\end{figure}

The line element (\ref{GeneralMetric}) for this solution  has a similar form as the usual Schwarzschild solution but with an additional term in the metric components $A(r)$, which can be expressed as follows:
\begin{equation}\label{SAdSMetric}
    A(r)=1-\frac{2M}{r}+\alpha^2r^2,
\end{equation}
and in this case $d\Omega^2=d\theta^2+\sin^2\theta d\varphi^2$ is the usual metric in the two-spheres  \cite{Socolovsky:2017nff}. By analyzing the function $A(r)$ we see that it exhibits a monotonically increasing behavior, tending to $-\infty$ as $r \rightarrow 0^+$ and to $\infty$ as $r \rightarrow \infty$. Moreover,  this metric owns a single event horizon given by the only real root of $A(r_H) = 0$. Fig.~\ref{Fig3} illustrates the behavior of $A(r)$, which diverges as $r\rightarrow\infty$, contrary to the usual Schwarzschild space-time.

As shown in the next section, there is a clear similarity between this metric and that of the black string, since the only term in $A(r)$ that differs from the corresponding term in the black string is the unit constant; the other terms have the same dependence. Note that the presence of the cosmological constant is not required for the existence of the solution, since $\alpha = 0$ yields the usual Schwarzschild solution. As opposed to that, in the case of the black string, it is not possible to have a cylindrically symmetric and stable solution without a cosmological constant \cite{Gregory:2000gf}, a solution we consider next.

\subsection{Black strings}

Let us now consider the general metric for a static black string in $3+1$-dimensions \cite{Lemos:1994xp}. For such a space-time, the metric \eqref{GeneralMetric} has cylindrical symmetry, $d\Omega^2=d\varphi^2+\alpha^2dz^2$, whereas the metric component $A(r)$ is given by
\begin{equation}\label{ArNeu}
    A(r)=\alpha^2r^2-\frac{\beta}{\alpha r}\ ,
\end{equation}
which in the charged case it is enlarged to
\begin{equation}\label{ArChar}
    A(r)=\alpha^2r^2-\frac{\beta}{\alpha r}+\frac{\gamma^2}{\alpha^2r^2}\ .
\end{equation}
where $\beta$ is a constant associated with the mass density of the string and $\gamma$ is a constant that refers to the charge density of the black string. 

\begin{figure}[t!]
\centering
\includegraphics[scale=0.66]{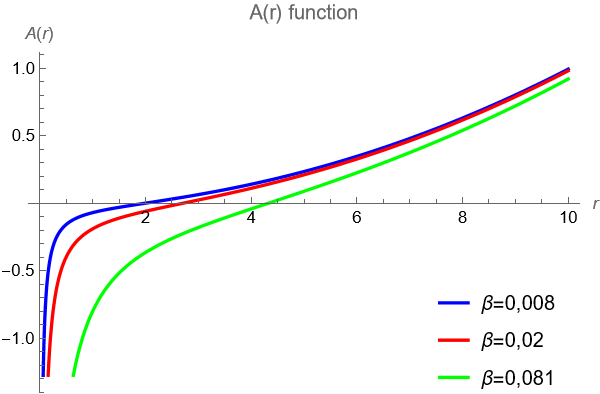}
\caption{The metric function $A(r)$ for different values of $\beta$ for the neutral black string \eqref{ArNeu} with $\alpha^2=0.1$.}
\label{Fig7}
\end{figure}

\begin{figure}[t!]
\centering
\includegraphics[scale=0.66]{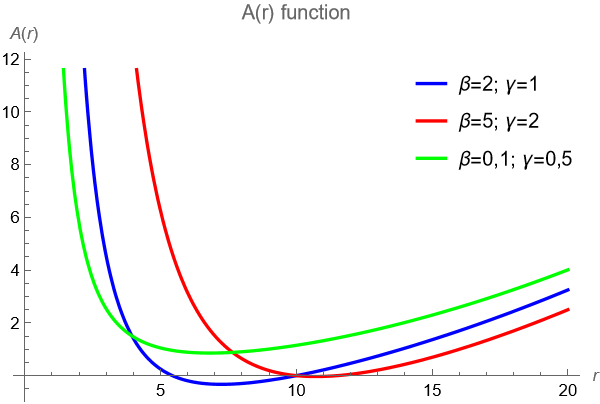}
\caption{The metric function $A(r)$ for different values of $\beta$ for the charged black string \eqref{ArChar} with $\alpha^2=0.1$.}
\label{FigBS}
\end{figure}

The neutral and charged string space-times above describe a $\mathcal{R}\times\mathcal{S}^1$ topology along the $z$ axis, approaching asymptotically the AdS space-time. As shown in \cite{Lemos:1994xp}, there are several cases that provide different descriptions of the overall structure of the space-time, depending on the relative values of the free parameters. For the simplest case of a neutral black string, the space-time exhibits an event horizon, while when the charge is not vanishing, one can have two horizons (one external event horizon and another internal one that represents a Cauchy horizon), a unique horizon in the extremal case, or a naked singularity in absence of any horizons. In Fig.~\ref{Fig7} the behavior of the function $A(r)$ is shown for the neutral case by considering several values of the parameters, and in Fig. \ref{FigBS} for the charged black string. As shown there, the metric is singular at $r=0$ for every case, corresponding to a true physical singularity is placed. The roots for $A(r)$ just represent different types of horizons, if any.

Our next goal is to perform a ray-tracing of the geodesic trajectories around these asymptotically AdS space-times, with a particular aim on the black string, prior to the generation of their optical appearances.

\section{Geodesics and Ray-Tracing}
\label{Geo}

\subsection{Ray-tracing procedure}

To track the light rays around the objects described by the space-time metrics shown in the previous section, let us study first the behavior of null geodesics is such space-times. The above metrics are spherically and cylindrically symmetric, respectively, and static, representing stationary black holes, which are described by the general form (\ref{GeneralMetric}), where $d\Omega^2 \equiv d\theta^2 + \sin^2\theta\, d\varphi^2$ for the spherically symmetric space-time, and $d\Omega^2 \equiv d\varphi^2 + \alpha^2 dz^2$ for the cylindrically symmetric case. Here, we are specifically interested in working in the equatorial plane of the space-time, that is, $\theta = \frac{\pi}{2}$ for the spherical case, and $z = 0$ for the cylindrical one. Thus, the line element in the equatorial plane, for both types of symmetry, is given by:
\begin{equation}\label{metric}
    ds^2=-A(r)dt^2+\frac{dr^2}{A(r)}+r^2d\varphi^2.
\end{equation}

To analyze the geodesics, we start with the geodesic equation,
\begin{equation}
    \frac{d^2x^{\mu}}{d\tau^2}+\Gamma^{\mu}{}_{\nu\lambda}\frac{dx^{\nu}}{d\tau}\frac{dx^{\lambda}}{d\tau}=0\ ,
\end{equation} 
where the curve is parametrised with the affine parameter $\tau$. Using the line element (\ref{metric}), the geodesic equation reads as
\begin{eqnarray}
    \label{3}
    \left(\frac{ds}{d\tau}\right)^2=-A(r)\left(\frac{dt}{d\tau}\right)^2+\frac{1}{A(r)}\left(\frac{dr}{d\tau}\right)^2\nonumber\\
    +r^2\left(\frac{d\varphi}{d\tau}\right)^2.
\end{eqnarray}
To simplify the notation, we will represent the derivatives with respect to the affine parameter with a dot on top of the function on which the derivative operator acts, such that $\frac{df}{d\tau}\equiv \dot{f}$. Moreover, we can state that $\left(\frac{ds}{d\tau}\right)^2\equiv \epsilon$, where $ds^2=0$ ($\epsilon=0$) for null trajectories and $ds^2=-d\tau^2$ ($\epsilon=-1$) for time-like geodesics. Then, Eq.(\ref{3}) can be rewritten as:
\begin{equation}\label{4}
    \epsilon=-A(r)\dot{t}^2+\frac{\dot{r}^2}{A(r)}+r^2\dot{\varphi}^2\ .
\end{equation}
In order to simplify the equations, the conserved quantities associated to the corresponding Killing vectors, time-translations and rotations along $\varphi$, can be easily obtained as
\begin{equation}\label{5}
    E=-g_{0\mu}\dot{x}^{\mu}=A(r)\dot{t} \quad ; \quad L=g_{2\mu}\dot{x}^{\mu}=r^2\dot{\varphi},
\end{equation}
where $E$ and $L$ are the constants of motion. Note that, since the space-time is not  asymptotically flat, we cannot state that $E$ e $L$ are the energy and the angular momentum per unit mass, as in usual asymptotically flat space-times. By substituting (\ref{5}) in (\ref{4}), one obtains the following expression:
\begin{eqnarray}\label{6}
    \epsilon&=&\frac{1}{A(r)}(-E^2+\dot{r}^2)+\frac{L^2}{r^2}\nonumber\\
    &\rightarrow& \dot{r}^2= E^2-V_{eff}(r)\ .\label{RadialEq1}
\end{eqnarray}
where we have defined the effective potential
\begin{equation}\label{Potential1}
    V_{eff}(r)=A(r)\left(\frac{L^2}{r^2}-\epsilon\right)\ .
\end{equation}
For the case of null geodesics (the case we are interested in here), $\epsilon=0$, and then Eq.\eqref{Potential1} leads to:
\begin{equation}\label{PotentialNull1}
     V_{eff}(r)=A(r)\frac{L^2}{r^2}\ .
\end{equation}
Circular orbits for photons, if exist, hold the conditions: $E^2=V_{eff}(r_c)$ and $V'_{eff}(r_c)=0$ (with primes denoting derivatives with respect to the radial variable), where $r_c$ is the radius of the circular orbit. There are two possible types of circular orbits: a stable one, which satisfies the condition $V''_{eff}(r_c) > 0$, and an unstable one, for which $V''_{eff}(r_c) < 0$. Black holes necessarily have at least one unstable circular orbit \cite{Carballo-Rubio:2024uas}, while for horizonless compact objects they always come (if present) in pairs, i.e., a stable and an unstable one \cite{Cunha:2017qtt}.

The next step to build the ray-tracing procedure for null geodesics is to introduce the impact parameter as defined 
by $b\equiv \frac{L}{E}$. Then, the equation for the radial motion \eqref{RadialEq1} in the $z=0$ plane yields:
\begin{equation}
    \frac{\dot{r}^2}{L^2}=\frac{1}{b^2}-\frac{A(r)}{r^2}\ ,
\end{equation}
where the affine parameter can be redefined in order to absorb the constant  $L^2$. Moreover, here we are interested in tracing the photons trajectories for $\varphi(r)$. Since $\dot{r}=\frac{dr}{d\varphi}\dot{\varphi}$, the following equation for such trajectories is obtained:
\begin{eqnarray} \label{eq:defang}
    \frac{d\varphi}{dr}=\pm \frac{b}{r^2\sqrt{1-\frac{b^2A(r)}{r^2}}}\ .
\end{eqnarray}
This equation provides the deflection angle $\varphi$ as a function of the impact parameter. For a given value of $E^2$ carried by the photon, the equation $E^2=V_{eff}(r)$ will be satisfied at certain $r=r_{0}$ such that $\dot{r}=0$, which means that $\varphi'(r)$ changes sign and the coordinate $r$ changes its behavior from decreasing (considering that the light ray is launched from infinity toward the origin) to increasing, therefore corresponding to a turning point. We can determine the relation between the corresponding impact parameter and such a radius $r_0$ from Eq.\eqref{RadialEq1} as
\begin{equation}
    b_0=\frac{r_0}{\sqrt{A(r_0)}}.
\end{equation}
Furthermore, whenever such a turning point is identified with the maximum of the effective potential, then the deflection angle provided by (\ref{eq:defang}) diverges. The corresponding surface $r=r_c$ at which this happens is dubbed as the {\it photon sphere} and photons asymptote to it whenever they have a {\it critical impact parameter} given by $b_c=\frac{r_c}{\sqrt{A(r_c)}}$. A photon sufficiently close to the photon sphere will be able to circle the black hole a number $n$ of half-turns and will create, on the observer's plane image, a number of {\it photon rings} indexed by the number $n$ \cite{Gralla:2019xty,Johnson:2019ljv,Vincent:2022fwj} and which approach, in the limit $n \to \infty$, a theoretical curve there, corresponding to the projection of the photon sphere.  

The critical curve therefore splits trajectories into three main groups. Those with $E^2<V_{eff}$ ($b>b_c$) will find a scattering point on their trajectory returning back to the observer and create a bright region there (dubbed as the direct image); those with $E^2>V_{eff}(r)$ ($b<b_c$) will find the black hole event horizon unimpeded, creating the typical shadow in the image; and finally those which asymptote to the photon sphere or hover near to it, $E^2 \lesssim V_{eff}(r)$ ($b \gtrsim b_c$), will create the sequence of secondary images, i.e., the photon rings. Such photon rings can be employed as tests of the Kerr metric and its alternatives, see e.g. \cite{Wielgus:2021peu}.

\begin{figure}[t!]
\centering
\includegraphics[scale=0.66]{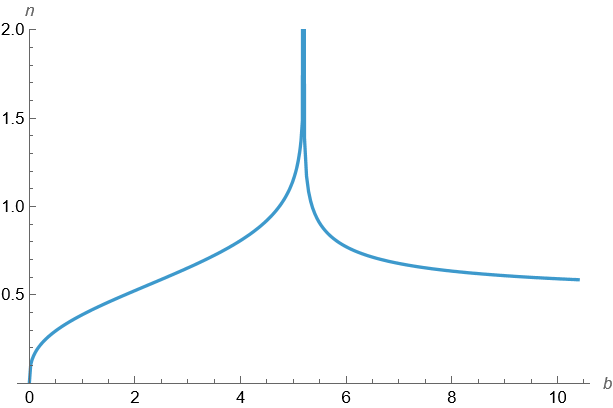}
\caption{Number of half- turns $n$ around an asymptotically flat Schwarzschild black hole as a function of the impact parameter $b$.}
\label{Fig1}
\end{figure}

\begin{figure}[t!]
\centering
\includegraphics[scale=0.66]{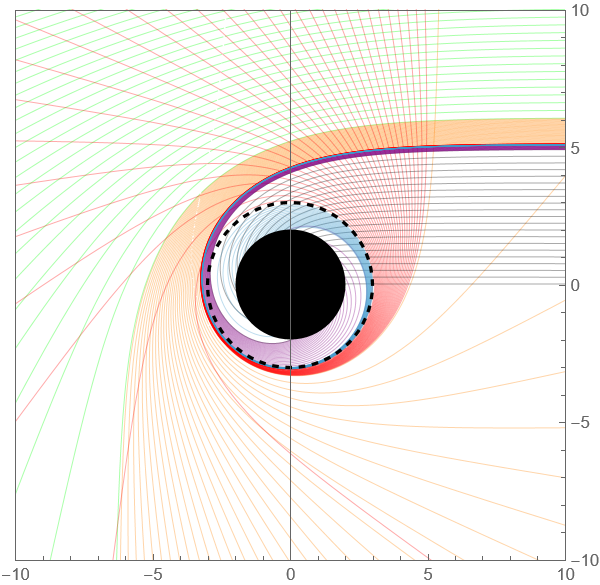}
\caption{Trajectories in the equatorial plane of light rays with $b \in (0,10)$ for the asymptotically flat Schwarzschild black hole. Colours correspond to the direct emission $n=0$ (green) and to the $n=1$ (orange) and $n=2$ (red) photon ring emissions in the observer's plane image. Those curves in gray, purple and light blue also turn $n=0,1,2$, respectively, but from the event horizon towards the asymptotic observer, therefore belonging to the shadow in the observer's image.}
\label{Fig2}
\end{figure}

Fig.~\ref{Fig1} shows the number of half-turns made by a photon as a function of the impact parameter for the case of the standard (asymptotically flat) Schwarzschild space-time. Furthermore, the ray-tracing of photons with impact parameters $b \in (0,10)$ are displayed  in Fig.~\ref{Fig2}. In this plot, the green lines represent photons undergoing direct emission (i.e. $n=0$), the orange lines represent $n=1$ photon ring emissions, while the red lines correspond to $n=2$ photon ring emissions.  The gray, purple and light blue lines represent, respectively, the trajectories of light rays that turn $n=0,1,2$ half-times from the event horizon towards the observer and which belong, in this ray-tracing procedure, to the black hole shadow region.

\subsection{Ray-tracing in Schwarzschild-AdS spacetime}
\begin{figure}[t!]
\centering
\includegraphics[scale=0.66]{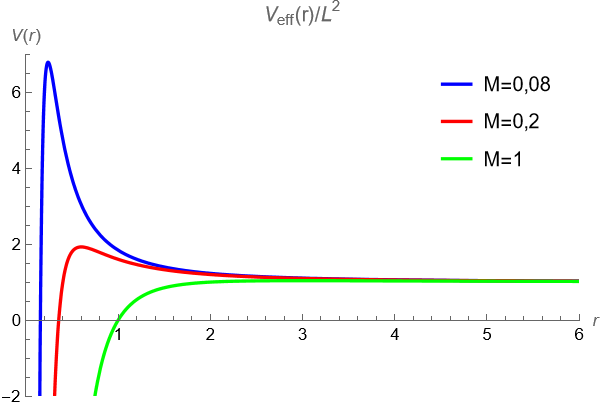}
\caption{Effective potential for null geodesics in Schwarzschild-AdS space-time for $\alpha=0.1$.}
\label{Fig4}
\end{figure}

For the metric defined in \eqref{SAdSMetric}, the effective potential for null geodesics is given by:
\begin{equation}\label{VeffSAdS}
    V_{eff}(r)=\left(1-\frac{2M}{r}\right)\frac{L^2}{r^2}+\alpha^2L^2\ .
\end{equation}
We see that the right-hand side of Eq.\eqref{VeffSAdS} is identical to that of the Schwarzschild case except for the additional constant term $\alpha^2L^2$. Hence, since the analysis of circular orbits depends on $V'_{eff}(r)$, this means that there is an unstable circular orbit at the same location of the asymptotically flat Schwarzschild case, namely, $r_c = 3M$. However, the non-asymptotically flat character of the modified space-time means that  the effective potential does not decrease towards infinity but tends asymptotically to a constant value $V_{eff}(r \rightarrow \infty) = \alpha^2 L^2$. This has the important consequence that the constraint $E^2 \geq \alpha^2 L^2$ for a well-defined geodesic equation implies a maximum value for the impact parameter $b$, which corresponds to the minimum energy of the photon, i.e., 
\begin{equation} \label{eq:upper}
\frac{L^2}{E^2} \leq \frac{1}{\alpha^2} \rightarrow b \leq \frac{1}{\alpha}
\end{equation}
This fact is illustrated via Fig.~\ref{Fig4}, where the non-vanishing asymptotics of the effective potential is clearly seen. The ray tracing of the geodesics follows a very similar procedure as the one of the usual Schwarzschild case, where we can have photon scattering when $b_c < b < \frac{1}{\alpha}$, an unstable circular orbit when $b = b_c$, and photon capture when $0 \leq b < b_c$. The main difference lies in the presence of the upper limit for the impact parameter given by (\ref{eq:upper}), which is a consequence of the AdS-type geometry. The critical impact parameter separating capture versus turning orbits is given in this case by
\begin{equation}
    b_c=\frac{3\sqrt{3}M}{\sqrt{1+27\alpha^2M^2}}
    \label{criticalbAdS}
\end{equation}

\begin{figure}[t!]
\centering
\includegraphics[scale=0.66]{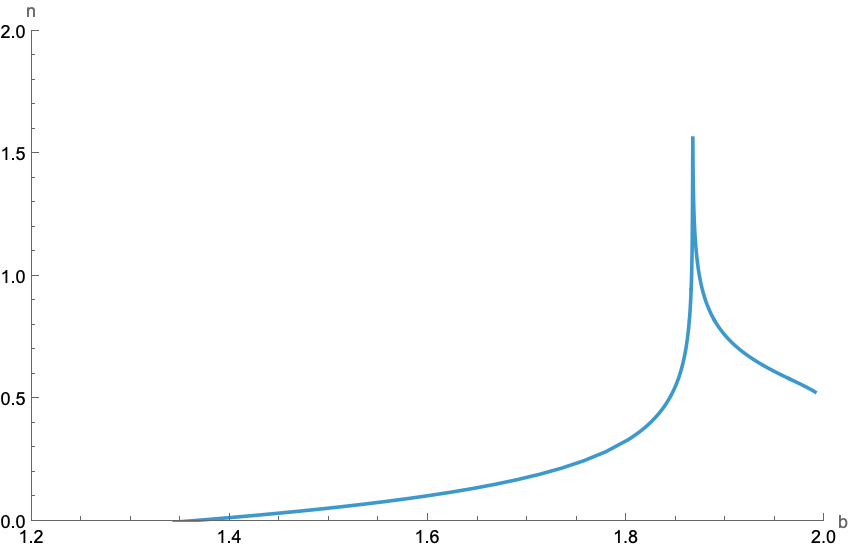}
\caption{Number of half- turns $n$ around an Schwarzschild-AdS black hole as a function of the impact parameter $b$.}
\label{Fig5}
\end{figure}

\begin{figure}[t!]
\centering
\includegraphics[width=8.6cm,height=8cm]{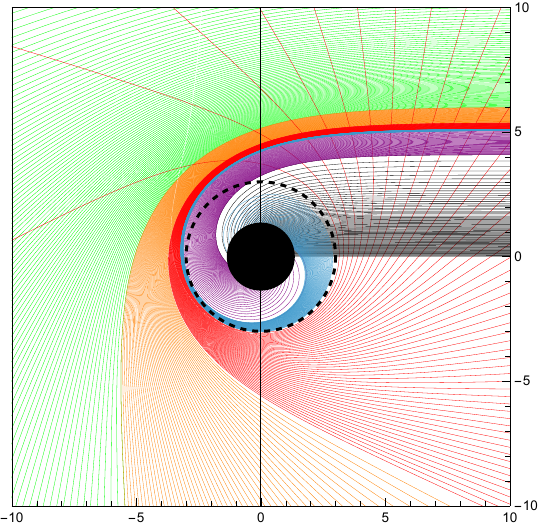}
\caption{Trajectories in the equatorial plane of light rays with $b \in (0,10)$ for the Schwarzschild-AdS black hole.}
\label{Fig6}
\end{figure}

In Fig.~\ref{Fig5} we depict  the number of half-orbits completed by the photons around the black hole as a function of the impact parameter. While it resembles the same behavior observed for the asymptotically flat Schwarzschild case discussed above (recall Fig.~\ref{Fig1}) in that the number $n$ increases as $b$ approaches the critical parameter, for $b>b_c$ it decreases until the impact parameter reaches its maximum value. Similarly,  Fig.~\ref{Fig6} displays the ray-tracing procedure for $b \in (0,10)$, with a similar distribution of colours for orbits as in Fig.~\ref{Fig2}. 


\subsection{Ray-tracing of a black string}
\begin{figure}[t!]
\centering
\includegraphics[scale=0.66]{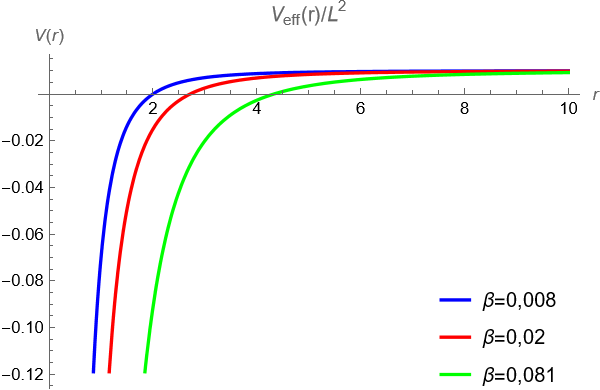}
\caption{Effective potential (\ref{PotentialNUll2}) for null geodesics around a neutral black string for $\alpha=0.1$ and different values of $\beta$.}
\label{Fig8}
\end{figure}

Let us apply here the procedure described above for computing and analyzing null geodesics of a black string. For the case of a neutral black string, the effective potential \eqref{PotentialNull1} reads as
\begin{equation}\label{PotentialNUll2}
    V_{eff}(r)=\left(\alpha^2r^2-\frac{\beta}{\alpha r}\right)\frac{L^2}{r^2}=\alpha^2L^2-\frac{\beta L^2}{\alpha r^3}\ .
\end{equation}
It is straightforward to show that the potential  \eqref{PotentialNUll2} has neither  maxima nor minima and, consequently, there does not exist any photon circular orbits for a neutral black string. This is easily seen in Fig.~\ref{Fig8}, where the potential \eqref{PotentialNUll2}  is depicted, such that the  absence of a maximum also makes the function not to tend asymptotically to zero for $r\rightarrow\infty$, as it could expected since the space-time is not asymptotically flat but AdS.

\begin{figure}[t!]
\centering
\includegraphics[scale=0.66]{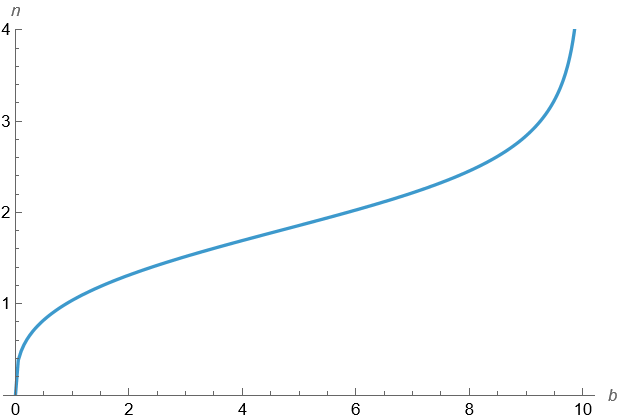}
\caption{Number of half-turns $n$ around a neutral black string, characterized by the effective potential (\ref{PotentialNUll2}), as a function of the impact parameter $b$.}
\label{Fig9}
\end{figure}

\begin{figure}[t!]
\centering
\includegraphics[scale=0.66]{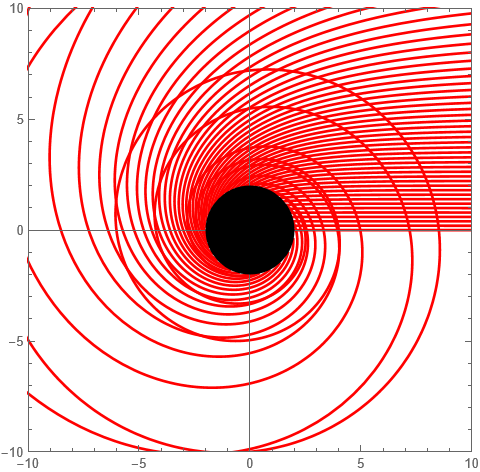}
\caption{Trajectories for photons around a neutral black string.}
\label{Fig10}
\end{figure}

For this neutral black string, the location of the event horizon is easily obtained as
\begin{equation}
    r_h=\frac{\beta^{1/3}}{\alpha}\ .
\end{equation}
allowing to express the function  $A(r)$ in \eqref{ArNeu} as
\begin{equation}\label{Ar2}
    A(r)=\alpha^2r^2\left(1-\frac{r_h^3}{r^3}\right)\ ,
\end{equation}
Hence, the metric just depends on two parameters, $\alpha$ and $r_h$, which are related to the cosmological constant and to the mass of the string, respectively. The equation for the trajectory is therefore written as 
\begin{equation}\label{23}
    \frac{d\varphi}{dr}=\pm\frac{b}{r^2\sqrt{1-b^2\alpha^2\left(1-\frac{r_h^3}{r^3}\right)}}.
\end{equation}
This expression is next used to compute the light trajectories. This way we follow the usual procedure of integrating Eq.(\ref{23}) for the neutral black string for a bunch of light rays corresponding to different values of $b$. We find a qualitatively similar behaviour as the one found for the Schwarzschild-AdS metric, in which the effective potential tends to a nonzero value in the asymptotic limit, which implies the existence of a maximum impact parameter given by $b = 1/\alpha$, corresponding to the constraint $E^2\geq \alpha^2L^2$. This means that, contrary to what happens in asymptotically flat space-times, every photon necessarily enters the horizon of the neutral black string, as expected for an asymptotically AdS space-time. Moreover, the larger the impact parameter of the photon is, the larger the number of turns around the black string becomes, approaching infinity for $b=b_{\text{max}}$. In Fig.~\ref{Fig9}, the number of half-orbits around the black string is depicted whereas Fig.~\ref{Fig10} shows a set of trajectories. Both figures display the main features of light trajectories, regarding the absence of a photon sphere and the presence of a maximum impact parameter, that is, trajectories approaching the latter have a larger number of half-turns around the black hole string than their counterparts with lower $b$ and every light trajectory eventually hits the black hole event horizon. This is neatly different from the ray-tracing associated to both asymptotically flat and AdS Schwarzschild black holes, which should have a reflection on the corresponding optical appearances.

For completeness of our analysis, let us consider now the case of a charged black string, whose effective potential for null trajectories reads as
\begin{equation}
    V_{eff}(r)=\left(\alpha^2r^2-\frac{\beta}{\alpha r}+\frac{\gamma^2}{\alpha^2r^2}\right)\frac{L^2}{r^2}=\alpha^2L^2-\frac{\beta L^2}{\alpha r^3}+\frac{\gamma^2L^2}{\alpha^2r^4}\ ,
    \label{PotChargeBS}
\end{equation}
Then, by solving $V'_{eff}(r)=0$, one finds a positive real root at:
\begin{equation}\label{14}
    r_c=\frac{4\gamma^2}{3\alpha \beta}.
\end{equation}
so circular orbits do exist in this case. In order to analyse their stability, the second derivative of the potential is obtained as
\begin{eqnarray}
    V''_{eff}(r_c)=\frac{4L^2\gamma^2}{\alpha^2r_c^6}>0.
\end{eqnarray}
Hence, since the second derivative of the effective potential with respect to the radial coordinate is positive, this circular orbit for photons is stable. In fact, the form of the effective potential \eqref{PotChargeBS} looks like the Kepler potential, where closed stable orbits exist. In addition, in the charged black string, the function $A(r)$ in \eqref{ArChar} may have up to two positive real roots, associated to two horizons, the external horizon $r_+$ and the internal one $r_-$, similar to what occurs in the Reisner-Nordstr\"om black hole, and furthermore extremal black hole configurations and naked singularities can also be present. In the black hole case the minimum of the effective potential, as given by (\ref{14}), is located between the two horizons,  which means that trajectories away from the horizon are similar to the neutral black string, since such circular photon trajectories are not allowed outside the external horizon.





\section{Optical appearance and shadows}
\label{MPIE}

\subsection{Transfer function}

To determine the optical appearance of a black hole or any other compact object, it is necessary to model the emission of the plasma that surrounds it according to several of its features, including absorptivity, scattering, and emission, among other effects that are known to play a heavy role such as the behaviour of magnetic fields and turbulence, requiring the development of sophisticated radiative transfer codes \cite{Gold:2020iql}. For the sake of this work we shall use a simplified modelling of the disk's emission, recognizing that a full account on this topic requires a more rigorous and complex analysis. In this sense, our goal is to understand the basics of how Schwarzschild-AdS black holes and black string images are formed and what features they display as compared to canonical black hole ones, avoiding the ``contamination" enacted by the disk as much as possible.

\begin{figure}[t!]
\centering
\includegraphics[scale=0.8]{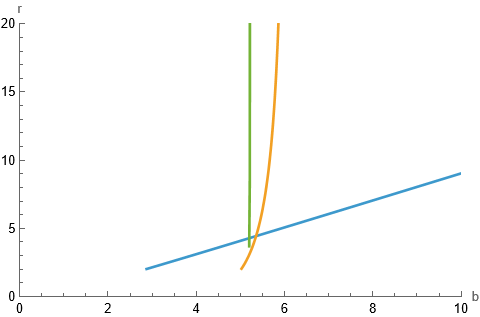}
\includegraphics[scale=0.8]{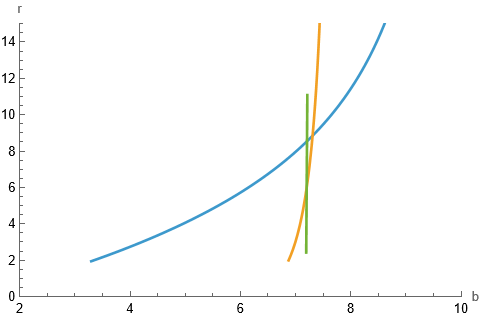}
\includegraphics[scale=0.58]{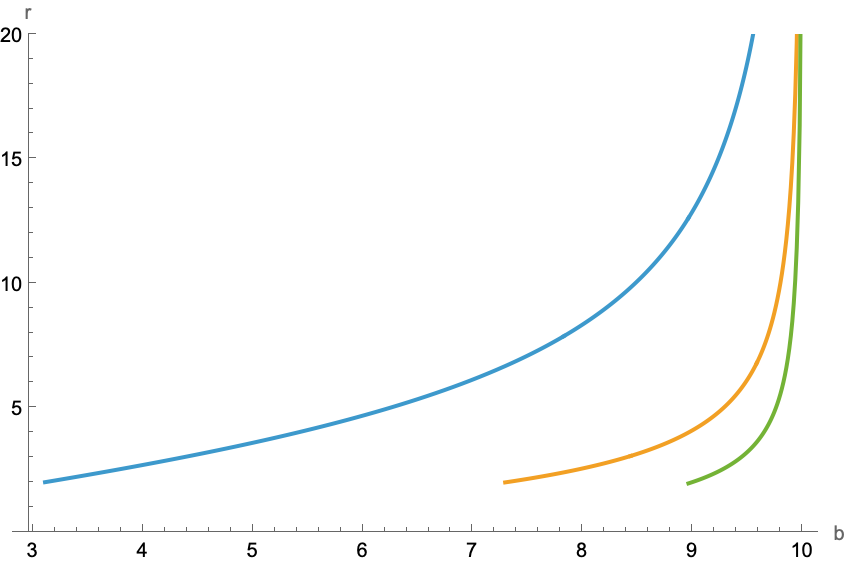}
\caption{Transfer function for (from top to the bottom): asymptotically-flat Schwarzschild, Schwarzschild-AdS and neutral black string. Colours correspond to the direct emission (blue), the $n=1$ photon ring (orange), and the $n=2$ photon ring (orange).}
\label{TransferF}
\end{figure}

In order to run the process of generation of images we first need to  identify the points in the accretion disk where photon emission occurs, as a function of the values  of the impact parameter for each type of emission, that is, the direct $n=0$ and photon ring $n=1$ and $n=2$ emissions. This provides the so-called transfer function, consisting of three different curves associated to each such emission. In Fig.~\ref{TransferF}  we depict  the transfer functions (from top to bottom) for the asymptotically flat Schwarzschild space-time, the Schwarzschild-AdS solution, and the black string, respectively.  On the vertical axis, we have the position where the geodesic intersects the accretion disk (assumed to be located on the equatorial plane), while on the horizontal axis we have the values of the impact parameter. In the canonical black hole space-time we see the three curves corresponding to the direct emission (blue), the $n=1$ photon ring (orange) and the $n=2$ photon ring (green), the latter two nearby the critical impact parameter $b=b_c$, given the fact that they are tightly associated to the corresponding photon sphere. Furthermore, it is known that the slope of such curves are directly connected to the degree of demagnification in the corresponding images of the direct and photon ring emissions \cite{Perlick:2021aok}.

The image for the Schwarzschild-AdS black hole displays the same three curves, but with some significant differences. In particular, the direct emission curve is neatly different than in the asymptotically flat Schwarzschild black hole, curving itself up as the impact parameter $b$ grows rather than having a constant slope. As for the photon ring emissions, there are also visual differences as compared to the usual Schwarzschild black hole such as being much more close to one another, which are expected to have a non-negligible impact in the corresponding images.

Things are very different for the black string. Indeed, now the direct emission region is entirely inside the photon ring emissions for every value of $b$, and the same happens with the $n=1$ photon ring as compared to the $n=2$ one. On the other hand, the slope of each curve grows very quickly as the maximum impact parameter value in this space-time is reached, which is in agreement with the results of the ray-tracing procedure and the behaviour of the number of half-turns $n$ displayed before. These features are expected to provide significant modifications to the corresponding images.

\subsection{Emission profiles} 

The next step in our setup is to establish the  particular luminosity profile of emission for the accretion disk. To this end, we shall first assume that the emission occurs in a optically thin disk with an emission region located in the equatorial plane only, and extending from a certain inner edge outwards \cite{Page:1974he}. We denote the emitted intensity at a given frequency $\nu$ by  $I^{em}_{\nu}$  while the  observed intensity at a frequency $\nu'$ is denoted by $I^{ob}_{\nu'}$. In our models we assume that the emitted intensity is isotropic and depends only on $r$, that is, $I^{em}_{\nu} = I_{\nu}(r)$. Furthermore, we neglect the effects associated with the emission and absorption coefficients present in the relativistic radiative transfer equation. As a consequence, the quantity $I_{\nu}/\nu^3$ remains invariant along the photon trajectory \cite{Mihalas:1984}. Since the observed frequency can be determined from the line element as $\nu' = \left(\frac{A(r)}{A(r_{ob})}\right)^{1/2}\nu$, where $r$ is the position of the emitted photon and $r_{ob}$ is the position of such an observed photon, we can use the invariance of $I_{\nu}/\nu^3$ to obtain the following relation between the observed and emitted luminosity:  
\begin{equation}
I^{ob}_{\nu'} = \left(\frac{A(r)}{A(r_{ob})}\right)^{3/2} I^{em}_{\nu}\ .
\end{equation} 
For asymptotically flat space-times we have simply $A(r_{ob})=1$; however,  since our space-time is asymptotically AdS then we cannot set $A(r_{ob})$ to one. To obtain the total intensity over all emitted frequencies, we integrate the expression for the observed intensity over each frequency, which results in
\begin{equation}
I^{ob} = \int I^{ob}_{\nu'}\, d\nu' = \left(\frac{A(r)}{A(r_{ob})}\right)^2 I(r)\ ,
\end{equation}
where we have defined $I(r) \equiv \int I^{em}_{\nu} \, d\nu$. 

Next, we incorporate the fact that light rays may cross the accretion disk several times, which would produce additional contributions to the brightness of the image. Therefore, adding intersections with the accretion disk up to the $n=2$ half-turn, we just need to sum up them as
\begin{equation}
    I^{obs}(b)=\frac{1}{A(r_{ob})^2}\sum_n [A^2I]\vert_{r=r_n(b)}\ ,
\end{equation}
where $r_n(b)$ is the transfer function analyzed above, and which allows to trade off radial dependences with impact parameter ones for every $n$-th intersection point of the photon trajectory with the accretion disk outside the event horizon. 

\begin{figure}[t!]
\centering
\includegraphics[scale=0.8]{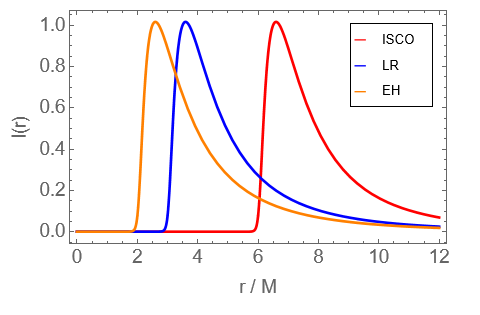}
\caption{Intensity profiles for different location of the peaks of emission, namely the ISCO, the photon sphere, and the event horizon in the SU distribution given by Eq.(\ref{IntensitySU}).}
\label{Fig11}
\end{figure}

In order to generate images, we assume the observer to be located very far away from the compact object, such that $A(r_{ob})\approx \alpha^2r_{ob}^2$, which is the term in the metric component that stands out in comparison to the others that become negligible for large $r_{ob}$. As for the intensity profile that describes the emission of the accretion disk, we make use  of some previously employed models in the literature (see e.g. \cite{Paugnat:2022qzy}), as given by suitable adaptations of  the Standard Unbound (SU) distribution, the later taking the form
\begin{equation}
    I^{em}_{SU}(r,\mu,\sigma,\gamma)=\frac{e^{-\frac{1}{2}\left[\gamma+\text{arc
    sinh}\left(\frac{r-\mu}{\sigma}\right)\right]^2}}{\sqrt{(r-\mu)^2+\sigma^2}},
        \label{IntensitySU}
\end{equation}
where the parameters $\mu$, $\sigma$, and $\gamma$ control the position of the emission peak, the scale (width), and the size of the central region, respectively. We choose these parameters so as for the locations of the  emission peak to correspond to the event horizon, the unstable circular photon orbit, and the innermost stable circular orbit for time-like observers (ISCO). In Fig.~\ref{Fig11} we display such profiles for  $I^{em}_{SU}$ as a function of $r$, finding significant differences in the effective region of emission between these three profiles.

\begin{figure}[t!]
\centering
\includegraphics[scale=0.57]{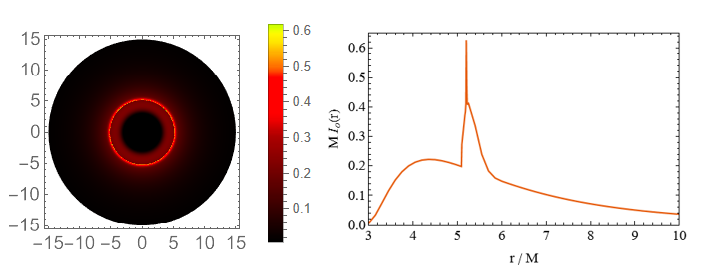}
\includegraphics[scale=0.57]{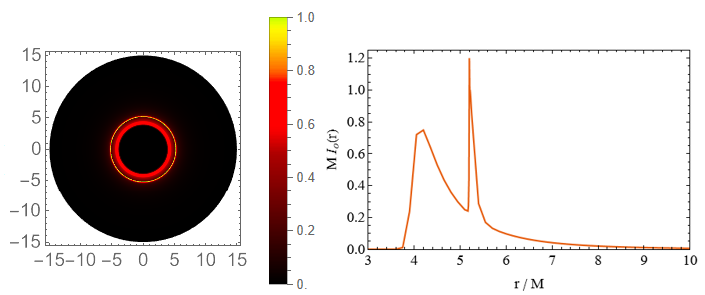}
\includegraphics[scale=0.57]{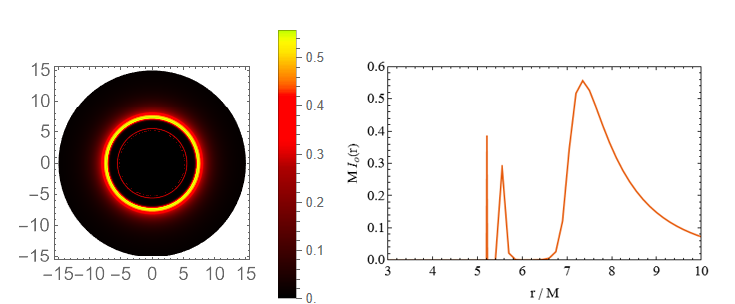}
\caption{Optical appearance (left figures) and observed intensity profiles (right figures) for an asymptotically flat Schwarzschild black hole. From top to bottom: intensity profile with the peak at the event horizon,  at the photon sphere,  and at the ISCO, respectively.}
\label{Fig12a}
\end{figure}

The corresponding optical appearances for these three models in the case of asymptotically flat Schwarzschild black holes is depicted in Fig. \ref{Fig12a}. There we neatly see the three contributions to the image as corresponding to the direct emission (which dominates all images) and the $n=1$ and $n=2$ photon rings. The latter appear as local insertions of luminosity that may appear overlapped with the direct emission (in the photon sphere and event horizon models) or separated from it (in the ISCO model). This is consistent with the expectations on these three classes of models as previously found in the literature (see also the discussion of \cite{Tsupko:2025hhf} on the overlapping of photon rings).

\begin{figure}[t!]
\centering
\includegraphics[scale=0.57]{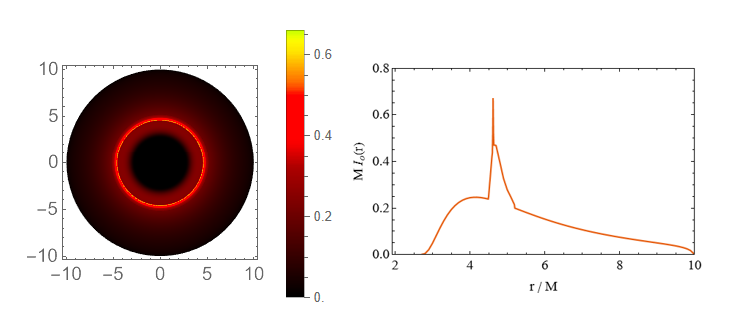}
\includegraphics[scale=0.57]{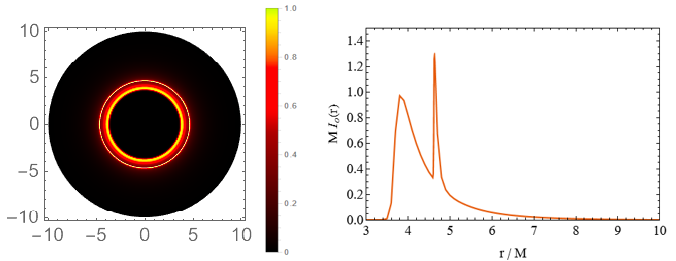}
\includegraphics[scale=0.57]{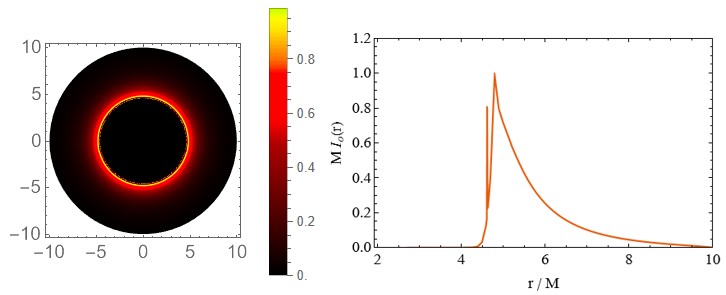}
\caption{Optical appearance (left figures) and observed intensity profiles (right figures) for the Schwarzschild-AdS black hole. From top to bottom: intensity profile with the peak at the event horizon,  at the photon sphere,  and at the ISCO, respectively.}
\label{Fig12}
\end{figure}

\subsection{Images in Schwarzschild-AdS spacetime}
\label{SAdSShadows}

For the Schwarzschild-AdS black hole, we take the metric (\ref{SAdSMetric}) with the values $M=1$ and $\alpha=0.1$, with the intensity profile $I^{em}_{SU}$ given in \eqref{IntensitySU} and the three surfaces displayed in  Fig.~\ref{Fig11}. The corresponding optical appearance generated for each case are shown in Fig.~\ref{Fig12}. We see once again the contributions of the direct and photon ring emissions, though now the way they are entwined with one another is clearly different as compared to the asymptotically-flat Schwarzschild configuration, in particular for the ISCO model in which photon rings are no longer separated from the direct emission, and which also changes the relative luminosity of the photon rings and the size of the central brightness depression are also apparent.

\begin{figure*}[t!]
\centering
\includegraphics[scale=0.75]{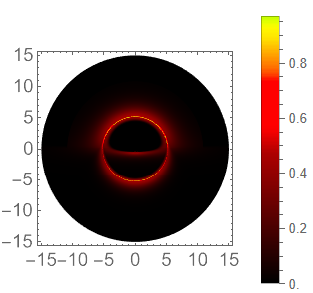}
\includegraphics[scale=0.75]{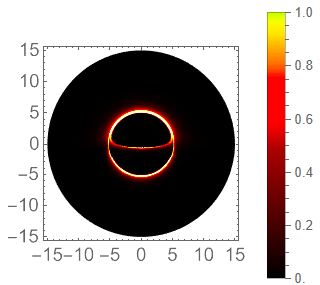}
\includegraphics[scale=0.75]{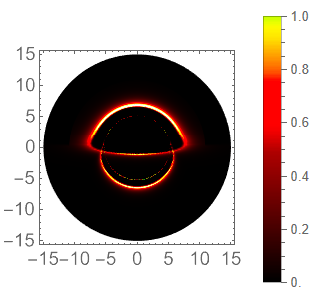}
\includegraphics[scale=0.75]{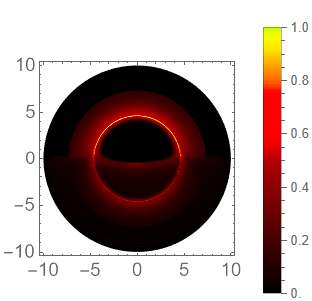}
\includegraphics[scale=0.75]{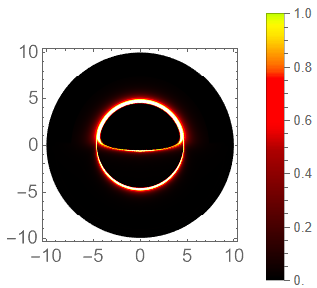}
\includegraphics[scale=0.75]{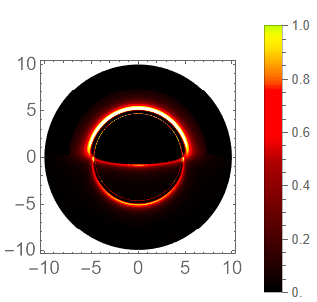}
\caption{Optical appearance for an asymptotically-flat Schwarzschild black hole (top figures) and the Schwarzschild-AdS black hole (bottom figures) where the accretion disk is inclined $80^{\circ}$. From left to right we display the images for the event horizon, photon sphere, and ISCO emission profiles, respectively.}
\label{Fig18a}
\end{figure*}

For both the asymptotically-flat Schwarzschild and the Schwarzschild-AdS black hole images, the accretion disk is located perpendicular to the line of sight of the observer. However, for more realistic observations, we will certainly not have a plane perfectly perpendicular to the observer's plane, so we must have an accretion disk tilted in relation to this frontal plane. By convention, we assume that the frontal plane has zero inclination, which increases when the plane tilts so that its upper part moves away from the observer.  In these cases, the same emission profiles are considered, but the simulations are adapted to account for the effect of the plane’s inclination. To this end, we consider an extreme inclination of $80^{\circ}$ in order to verify whether significant changes occur compared to the shadows of the frontal plane, and to see what the images of the inclined accretion disk reveal that differs from the non-inclined case. In the set of Figs. \ref{Fig18a} we show the result for the asymptotically-flat Schwarzschild solution (top figures) and for the Schwarzschild-AdS case (bottom figures) for the three SU models discussed above. By comparison of these two sets of images we see that they quite resemble one another, with the major difference being the size of the shadow and a more overlapped distribution of the photon rings with the direct emission in the AdS case than in the asymptotically-flat one.

\subsection{Images of black strings}

\begin{figure}[t!]
\centering
\includegraphics[scale=0.4]{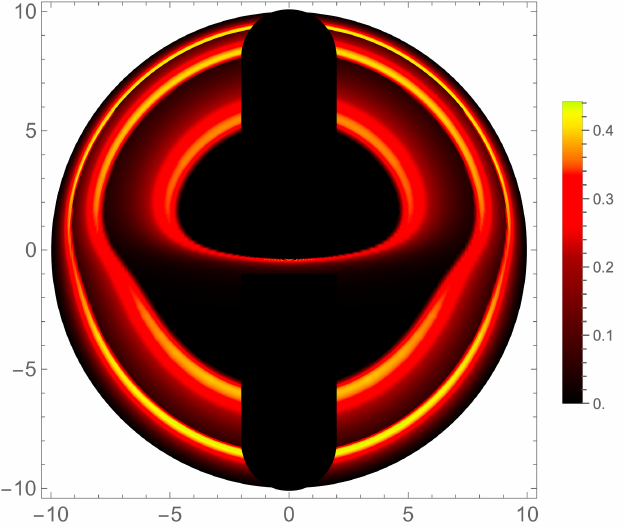}
\includegraphics[scale=0.4]{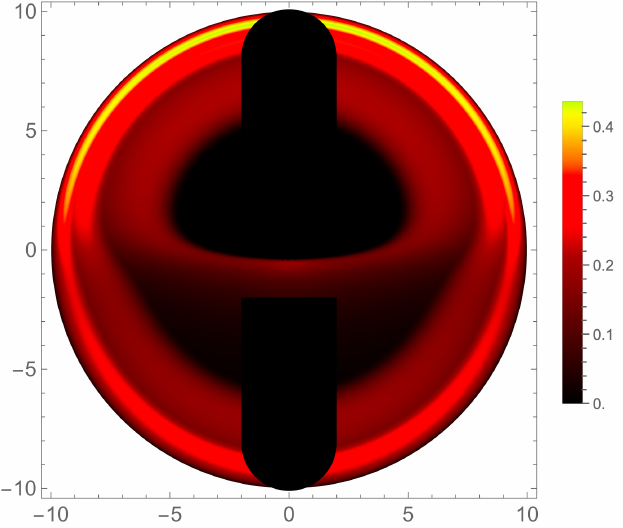}
\caption{Two samples of the optical appearance for a black string. Here the accretion disk and the black string are inclined $80^{\circ}$ and its peak of emission is located at the event horizon. Here, two sets of the free parameters for the emission profile (\ref{IntensitySU}) are considered: $\{\gamma=-1, \sigma=1/12\}$, $\{\gamma=-1, \sigma=1/4\}$, both of which peak at the event horizon. Here, the black tube represents the black string, which extends all along the $z$-axis.}
\label{BlackStringShadows}
\end{figure}

In the case of the black string, some precautions must be taken due to the new features of the space-time. Since we do not have spherical symmetry, it is not reasonable to assume that the accretion disk lies in a plane orthogonal to the equatorial plane containing the ``$z$'' axis, because in this case it would have to pass through the black string itself, as the string extends along the entire axis. The most logical assumption is therefore that the accretion disk surrounds the black string in the equatorial plane itself. Another point to take into account is that there is no circular photon orbit around the black string. It can also be verified that there are no circular orbits for massive particles either, meaning that there is no ISCO orbit for the metric defined in Eqs.(\ref{GeneralMetric}) and (\ref{ArNeu}). Therefore, in comparison to the above cases, the only possible position for the emission peak is at the event horizon itself. Fig.~\ref{BlackStringShadows} shows some examples of images of black strings generated with the accretion disk inclined at $80^{\circ}$. As depicted there, the black string extends perpendicularly to the accretion disk plane, such that some photons coming from behind are trapped by the string. In addition, in comparison to the Schwarzschild case, there is no any inner dimmer photon ring, as expected due to the absence of a circular orbit for photons. Furthermore, this is an object with far more wider luminous regions in the same space of impact parameters as in the previous objects, including regions with boosts of luminosity.   All together makes this object to strongly depart from the expected features of canonical black hole images.

%

\section{Conclusion}
\label{Conclusions}

Future expectations on very long baseline interferometry concerning the possibility of disentangling the contributions of the space-time geometry and the accretion disk, particularly via the analysis of the photon rings, have pushed forward the research and simulation of images from accretion disks for different types of objects. The ultimate aim is to test any putative deviation from the Kerr solution of GR. Within this context, we have explored in this paper the effects produced in the optical appearance when considering space-times in which the natural asymptotic flatness assumption is dropped in favour of an asymptotically Anti-de Siter one. This is motivated by the fact that AdS space-times are well studied and widely used in the community, largely due to the AdS/CFT correspondence, which provides a valuable framework for investigating aspects of quantum gravity.

Two types of stationary asymptotically AdS space-times were studied, one with spherical symmetry (Schwarzschild-AdS) and another with cylindrical symmetry, the so-called black string. By analyzing the null trajectories and performing the ray-tracing procedure, we find that the Schwarzschild-AdS black hole quite resembles its asymptotically flat counterpart, as shown by comparing Figs.~\ref{Fig2} and \ref{Fig4}. In this sense, the most relevant novelty is the fact that for the AdS black hole the effective potential does not go to zero at infinity due to the presence of a negative cosmological constant. This also slightly modifies the critical impact parameter, which affects not only the number of half-turns that a photon with a particular impact parameter makes around the black hole before reaching the observer's screen (Figs.~\ref{Fig1} and \ref{Fig5}) but also the transfer functions, as shown in the top panels of Fig.\ref{TransferF}, which determine the distance at which a photon is emitted by the accretion disk, and consequently has a direct effect on the direct and photon rings' observed luminosity, as can be easily noticed by comparing the shadows for both types of black holes depicted in Figs.~\ref{Fig12a} and \ref{Fig12} for a perpendicular position of the accretion disk and Fig.~\ref{Fig18a}  when considering some inclination for the disk. 

The case of the black string is quite more subtle. Indeed, black strings are not only non-asymptotically flat objects but also own cylindrical symmetry, where the string behaves as a black hole with its corresponding event horizon but with an infinite extension along the $z$-axis. This way, by assuming the black string to be perpendicular to the observer, the only reliable position for the accretion disk, according to the cylindrical symmetry of the black string and to the geometrically thin assumption on the disk, is to be quasi-perpendicular to the line of sight. In addition, the black string space-time does not contain  any circular orbits for photons but its structure imposes that as a photon approaches the maximum allowed impact parameter $b_{max}$ (its existence a consequence of the AdS space), it indefinitely increases its number of half-turns around the black string, as shown in Fig.~\ref{Fig9} and by the trajectories depicted in Fig.~\ref{Fig10}. As a consequence, the transfer functions grow quickly near the $b_{max}$ impact parameter for every type of emission, such that the distribution of the brightness structure around the black string turns out to be very different in comparison to the above cases, as depicted in Fig.~\ref{BlackStringShadows}. Those differences do not lie only on the presence of the black string in the image but also on the structure of of photon rings which are dominated basically by the direct emission and the luminosity profile of the disk  (since there is no unstable circular null orbit). Such new features make these objects very different in their cast images as  compared to canonical black hole objects. 

To conclude, while from current cosmological observations one does not expect a universe governed by a negative cosmological constant, and furthermore all these results on asymptotically AdS objects make them hardly compatible with current observations by the EHT, the techniques and results considered here can be used to track qualitative and quantitative departures from canonical black holes from new compact structures in asymptotically non-flat space-times. In fact, the combination of the latter with other symmetries like the cylindrical one of the black string considered in this work provides quite strong deviations from the Kerr paradigm. This work, therefore, contributes to the collective effort in looking for suitable observational discriminators of alternative ultra-compact objects to be searched for in future upgrades of VLBI techniques. Further work along this perspective is currently underway.

\section*{Acknowledgements}

This work is supported by the Spanish National Grants PID2022-138607NBI00, PID2024-157196NB-I00, and CNS2024-154444 grants, funded by MICIU/AEI/10.13039/501100011033; the Q-CAYLE project, funded by the European Union-Next Generation UE/MICIU/Plan de Recuperacion, 
Transformacion y Resiliencia/Junta de Castilla y Le\'{o}n (PRTRC17.11) and the financial support by the Department of Education, Junta de Castilla y Le\'{o}n and FEDER Funds, 
Ref.~CLU-2023-1-05.

 
\end{document}